\documentclass[12pt,oneside,english]{amsart}
\usepackage[LGR,T1]{fontenc}
\usepackage[latin9]{inputenc}
\usepackage{babel}
\usepackage{amsthm}
\usepackage{graphicx}
\usepackage{setspace}
\usepackage[unicode=true,
 bookmarks=true,bookmarksnumbered=false,bookmarksopen=false,
 breaklinks=false,pdfborder={0 0 1},backref=false,colorlinks=false]
 {hyperref}
\hypersetup{pdftitle={New approaches in Math Demography},
 pdfauthor={Rao and Carey}}
\usepackage{breakurl}

\makeatletter

\DeclareRobustCommand{\greektext}{%
  \fontencoding{LGR}\selectfont\def\encodingdefault{LGR}}
\DeclareRobustCommand{\textgreek}[1]{\leavevmode{\greektext #1}}
\ProvideTextCommand{\~}{LGR}[1]{\char126#1}

\providecommand{\tabularnewline}{\\}

\theoremstyle{plain}
\newtheorem{thm}{\protect\theoremname}
\theoremstyle{remark}
\newtheorem{rem}[thm]{\protect\remarkname}

\usepackage{lscape}
\usepackage{lscape}
\usepackage{lineno}

\makeatother

\providecommand{\remarkname}{Remark}
\providecommand{\theoremname}{Theorem}

\begin{document}

\title{Computation of life expectancy from incomplete data}

\maketitle
\begin{center}

\textbf{\large{}Arni S.R. Srinivasa Rao}\footnote{\textbf{\large{}Corresponding author}}{\large\par}

Augusta University

1120 15th Street

Augusta, GA 30912, USA

Email: arrao@augusta.edu

\end{center}

\vspace{1cm}

\begin{center}

\textbf{\large{}James R. Carey}{\large\par}

Department of Entomology

University of California, 

Davis, CA 95616 USA 

and 

Center for the Economics and Demography of Aging

University of California, Berkeley, CA 94720

Email: jrcarey@ucdavis.edu

\end{center}

\vspace{1cm}
\begin{abstract}
Estimating the human longevity and computing of life expectancy are
central to the population dynamics. These aspects were studied seriously
by scientists since fifteenth century, including renowned astronomer
Edmund Halley. From basic principles of population dynamics, we propose
a method to compute life expectancy from incomplete data. 
\end{abstract}



\section*{\textbf{Introduction}}

In 1570 the Italian mathematician Girolamo Cardano suggested that
a man who took care of himself would have a certain life expectancy
of $\alpha$ (so that at any age $x$ we could expect him to live
$e(x)=\alpha-x$ more years) and then asked how many years would be
squandered by imprudent lifestyles \cite{Smith=000026Keyfitz}. Cardano's
healthiest man might be born with the potential of living to 260 years
but die at age 80, having wasted away years due to bad habits and
other such ill-advised choices. In this work, Cardano was in good
company; mathematicians such as Fibonacci, d\textquoteright Alembert,
Daniel Bernoulli, Euler, Halley, Lotka and many others contributed
to our understanding of population dynamics through mathematical models.
We can trace the notion of life expectancy in particular back to the
seventeenth century astronomer Edmund Halley who developed a method
to compute life expectancy \cite{Halley}. His studies led him to
observe \textquotedbl how unjustly we repine at the shortness of
our Lives, and think our selves wronged if we attain not Old Age;\textquotedbl{}
for \textquotedbl one half of those that are born are dead in Seventeen
years time\textquotedbl{} and to urge readers \textquotedbl that
instead of murmuring at what we call an untimely Death, we ought with
Patience and unconcern to submit to that Dissolution which is the
necessary Condition of our perishable Materials, and of our nice and
frail Structure and Composition: And to account it as a Blessing that
we have survived, perhaps by many Years, that Period of Life, whereat
the one half of the whole Race of Mankind does not arrive\textquotedblright{}
{[}postscript to \cite{Halley}). Besides his philosophical musings
Halley's essay contained many tables and detailed analyses.

Life expectancy at birth, is defined as the number of years remaining
to the average newborn. It is arguably the most important summary
metric in the life table because it is based on and thus reflects
the longevity outcome of the mortality experience of newborns throughout
their life course. When life expectancy is referred to without qualification
the value at birth is normally assumed \cite{Preston}. Life expectancy
is intuitive and thus easily understandable by lay persons, independent
of population age structure, an indicator of health conditions in
different societies, used in insurance annuity computations and as
a baseline for estimating the impact on longevity of diseases (e.g.
AIDS; cancer, diabetes) and lifestyle choices (e.g. smoking; alcohol
consumption). The value of life expectancy at birth is identical to
the average age in a life table population. The difference in life
expectancies between men and women is known as the gender gap. The
inverse of life expectancy equals both the per capita birth $(b)$
and per capita death $(d)$ rates in stationary populations( $b-d=0)$.
And since $b+d$ is a measure of the number of vital events in a population,
double the inverse of life expectancy equals what is referred to as
\textquotedblleft population metabolism\textquotedblright{} as applied
to stationary populations. Life expectancy at birth is the most frequently-used
comparative metric in biological studies of plants and animals. 

The first substantive demographic work in which life expectancy was
estimated was the \textquotedbl Bills of Mortality\textquotedbl{}
published in 1662 by John Graunt \cite{Graunt} who noted \textquotedbl From
when it follows, that of the said 100 conceived there remains at six
years 64, at thirty-six 26 at sixty-six 3 and at eighty 0\textquotedbl .
Although Edmund Halley \cite{Halley} and Joshua Milne \cite{Milne}
both introduced life table methods for computing life expectancy,
George King \cite{King} is generally attributed to introducing the
life table and life expectancy in modern notation. It was not until
1947 that life tables in general and life expectancy in particular
were introduced to the population biology literature for studying
longevity in non-human species \cite{Deevey}. Although life expectancy
is computed straightforwardly from life table survival data, complete
information is often not available. 

Therefore our objective in this paper is to describe a model that
we derived for use in estimating life expectancy at birth from a limited
amount of information. The information required to estimate life expectancy
in a given year with our model includes the number of births, the
number of infant deaths, and the number in the population at each
age from 0 through the maximal age, \textgreek{w}. Our computational
concept for $\omega=2$ is based on the following logic: (1) person-years
lived for a newborn cohort during the first year is the difference
between the number born and the number of infants that died. \emph{Person-years}
is the sum of the number of years lived by all persons in a cohort
or population. The number of person-years equals the life expectancy
of this cohort if their maximal age is one year (i.e. $l(1)$ = maximal
age); (2) person-years lived for this cohort during their first two
years of life is equal to person years lived up to one year and person
years that would be lived by people who have lived up to age 1. Person-years
lived by the newborn cohort during their second year of life is less
than the person years lived by newborn during the first year; (3)
the hypothetical number of person-years lived by the newborn cohort
during their third year of life (i.e. $l(3)$ = 0) equals the number
in the birth cohort minus the person-years lost due to deaths during
the third year. We use number of newborn and population at age 1 to
compute person years to be lived by newborn during first three years
of life. And this process continues through the oldest age, \textbf{$\omega>2$. }

\begin{landscape}

\begin{figure}
\includegraphics[scale=0.70]{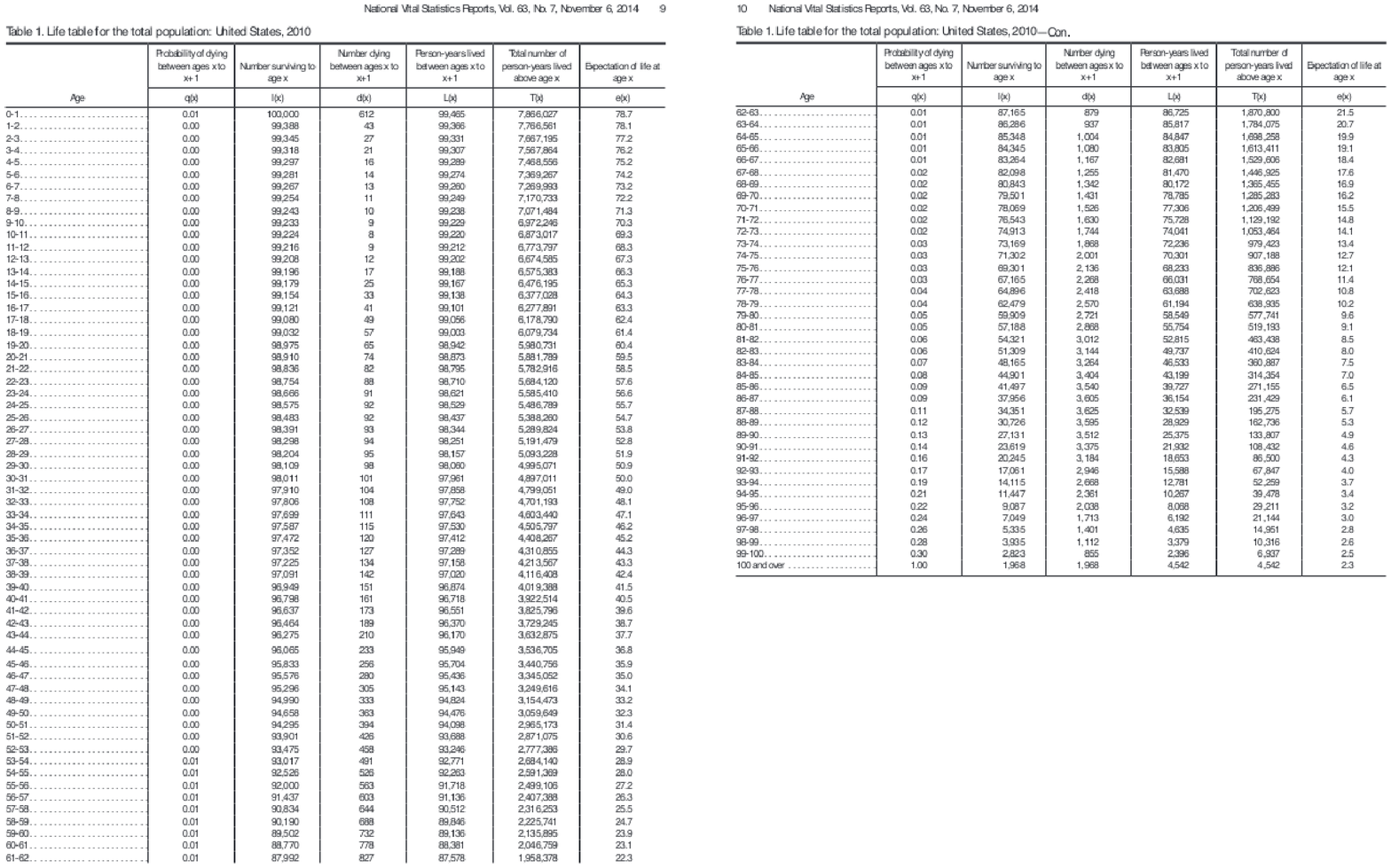}

\caption{\label{US Life Table}United States life table for the year 2010.
This life table was directly taken from National Vital Statistics
Reports \cite{Arias} }

\end{figure}

\end{landscape}

\begin{figure}
\includegraphics[scale=0.8]{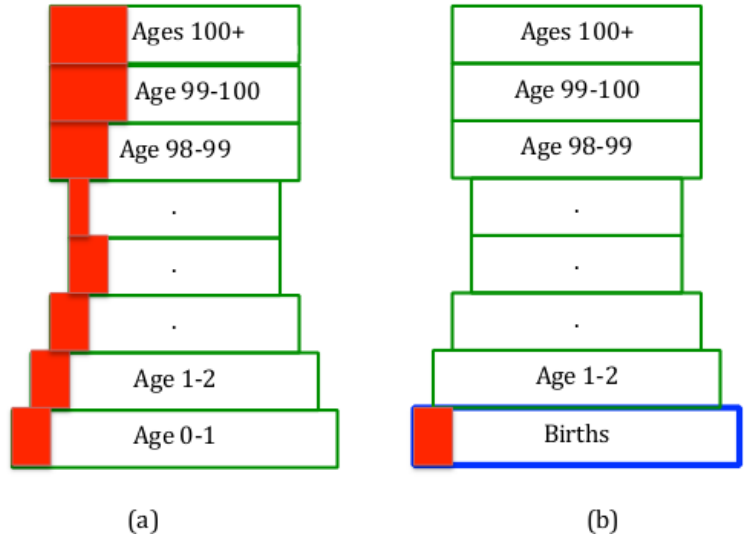}

\caption{\label{LT-DE Fig} (a) Data needed for life table approach. (b) Data
needed for computing life expectancy through new approach. Green bordered
rectangles are populations and red colored rectangles are death numbers
in the respective ages for an year. Blue-bordered rectangle is birth
numbers for an year. }

\end{figure}

Traditionally, the life expectancy of a population is computed through
life table techniques.\textbf{ }Life table of a population is a stationary
population mathematical model which primarily uses populations and
death numbers in all the single year ages for an year or for a period
of years to produce life expectancy through construction of several
columns. The last column of the life table usually consists of life
expectancies for each single year ages and first value of this column
is called life expectancy of the corresponding population for the
year for which the life table was constructed. See Figure \ref{US Life Table}
for the life table of US population in 2010 \cite{Arias}. There are
seven columns in this life table and the second column in the Figure
\ref{US Life Table}, which consists the values of probability of
dying\textbf{ }between ages $x$ to $x+1$ for $x=0,1,...100+$ is
first computed from the raw data (See Figure \ref{LT-DE Fig} for
the data needed for a life table) and other columns are derived from
the second column using formulae without any raw data. The last column
of the table in Figure \ref{LT-DE Fig} consists the values of expectation
of life at age $x$ for $x=0,1,...,100+.$ The first value in the
last column of the table in the Figure \ref{US Life Table} is 78.7,
which means life expectancy for the new born babies during 2010 in
the US population (boys and girls combined who are of aged 0-1 during
2010) is 78.7 years\textbf{. }In\textbf{ \cite{Arias} }we give the
various steps involved at Figure \textbf{\ref{US Life Table}}. 

For standard life table methods, see \cite{Keyfitz=000026Caswal},
for recent developments in computing life expectancy see \cite{Bon=000026FeenPNAS},
for astronomer Edmund Haley's life table constructed in 17th century,
see \textbf{\cite{Smith=000026Keyfitz}.} Recent advances in the theory
of stationary population models \cite{Rao =000026 Carey} are serving
the purpose of computing life expectancies for populations in the
captive cohorts \cite{MathDigest}. We propose a very simple formula
for computing life expectancy of newly born babies within a time interval
when age-specific death rates and life tables are not available. Age-specific
death rates at age $a$ are traditionally defined as the ratio of
the number of deaths at age $a$ to the population size at age $a$
\cite{Keyfitz=000026Caswal}. The method of calculating life expectancies
given in standard life tables uses age-specific death rates which
is computed from deaths and populations in each single year ages.
Refer to Figure \ref{LT-DE Fig} for the data needed in traditional
life table approach and for the newly proposed method. 

In this paper, we propose a formula for computing life expectancies
is comparable to the technique used to calculate life expectancies
in standard life tables, but can be applied when limited data is available.
The derived formula uses effective age-specific population sizes,
the number of infant deaths, and the number of live births within
a year. The number of infant deaths is usually defined as the number
of deaths within the first year of life in human populations. If the
study population is insects, necessary data can be considered within
any appropriate time interval. We tested our proposed simple formula
on both small hypothetical populations and global human populations.
When a sufficient amount of data on age-specific death rates is available,
the life table-based life expectancy is still recommended. 

\section*{\textbf{Life Expectancy of newly born babies }}

In this section we derive a formula for the life expectancy from basic
elements of population dynamics, namely, population-age structure
over two time points, simple birth and infant death numbers observed
over an interval of time. Suppose, the global population at the beginning
of times $t_{0}$ and $t_{1}$ (for $t_{0}<$$t_{1}$) is known, and
we are interested in finding the life expectancy of the people who
are born during $[t_{0},t_{1}).$ We assume the following information
to be known: i) $P(t_{0})$, the effective population size by single-ages
during $[t_{0},t_{1}$), which is indirectly computed as a weighted
or ordinary average of respective population sizes by single-ages
that are available at the beginning of $t_{0}$ and at the end of
$[t_{0},t_{1})$, ii) the number of live births, $B(t_{0})$, and
iii) the number of infant deaths, $D_{0}(t_{0})$ during the period
$[t_{0},t_{1})$. These quantities of known information are expressed
as,

\begin{eqnarray*}
P(t_{0}) & = & \int_{0}^{\omega}P_{i}(t_{0})di=\int_{0}^{\omega}\left[\frac{a_{i}P_{t_{0}}(i)+b_{i}P_{t_{1}}(i)}{a_{i}+b_{i}}\right]di\\
B(t_{0}) & = & \int_{t_{0}}^{t_{1}}B(s)ds\\
D_{0}(t_{0}) & = & \int_{t_{0}}^{t_{1}}D_{0}(s)ds
\end{eqnarray*}

where $P_{i}(t_{0})$ is the effective population aged $[i,i+1)$
for $i=0,1,...,\omega$ during $[t_{0},t_{1})$, with $P_{\omega}(t_{0})=0,$
for an age $\omega$ which is the next integer larger than the age
of eldest surviving person in $P(t_{0})$. $P_{t_{0}}(i)$ and $P_{t_{1}}(i)$
are observed populations in the age group $[i,i+1)$ at the beginning
of $t_{0}$ and at the end of $[t_{0},t_{1})$, $a_{i}$ and $b_{i}$
are population weights corresponding to $P_{t_{0}}(i)$ and $P_{t_{1}}(i)$,
respectively. $B(s)$ is the number of births at a given time $s\in[t_{0},t_{1})$
and $D_{0}(s)$ is the number of infant deaths for $s\in[t_{0},t_{1})$. 

We use standard life table notations to relate the quantities of the
population cohort life expectancy. Let, $l(x)$ be the number of survivors
of $B(t_{0})$ at age $x$ for $x=0,1,2,...,\omega.$ Clearly, $l(0)=B(t_{0})$
and $l(1)$ is approximated as, $l(1)\approx B(t_{0})-D(t_{0})$.
Suppose, $l(2)=0$. This implicitly implies that we have only observed
the data for $P_{0}(t_{0}),$ $P_{1}(t_{0})$, $B(t_{0}),$ $D_{0}(t_{0})$
during $[t_{0},t_{1})$. We will now use the concept of person-years,
which is a technical phrase in the life table model. Person-years
of a cohort represents the average future life time to be lived by
the cohort. The person-years lived by $B(t_{0})$ during their first
year of life (after removing person-years lost due to deaths), and
person-years lived by the remaining individuals of $B(t_{0})$ who
are surviving at age 1, (and removing deaths that occurred during
the second year of their life) and by assuming the deaths are uniformly
distributed over the age intervals $[0,1)$ and $[1,2)$ are:

\begin{eqnarray}
\int_{t_{0}}^{t_{1}}B(s)ds & - & \frac{1}{2}\int_{t_{0}}^{t_{1}}D_{0}(s)ds\label{eq:L0}
\end{eqnarray}

and

\begin{eqnarray}
\frac{1}{2}\int_{t_{0}}^{t_{1}}B(s)ds & - & \frac{1}{2}\int_{t_{0}}^{t_{1}}D_{0}(s)ds.\label{eq:L1 when l(2)=00003D0}
\end{eqnarray}

The total person-years that would be lived by $B(t_{0})$ during their
first two-years of life is

\begin{eqnarray}
\frac{3}{2}\int_{t_{0}}^{t_{1}}B(s)ds & - & \int_{t_{0}}^{t_{1}}D_{0}(s)ds.\label{eq:L1}
\end{eqnarray}

The life expectancy of $B(t_{0})$, i.e. new born babies at \textbf{$[t_{0},t_{1})$}
is,

\begin{eqnarray}
\frac{3}{2} & - & \frac{\int_{t_{0}}^{t_{1}}D_{0}(s)ds}{\int_{t_{0}}^{t_{1}}B(s)ds}.\label{LE-l(2)=00003D0}
\end{eqnarray}

When $P(t_{0})=\int_{0}^{3}P_{i}(t_{0})di$, then $l(1)\neq0$ and
$l(2)\neq0.$ The expression of $l(1)$ becomes $\int_{t_{0}}^{t_{1}}P_{1}(s)ds.$
We assume, $l(2)\approx2P_{1}(t_{0})-l(1)$, which implies the person-years
lived by $B(t_{0})$, during their second year of life is approximately
the same as the effective population at age $1$ during $[t_{0},t_{1})$,
instead of the previously obtained quantity in (\ref{eq:L1 when l(2)=00003D0})
(note that this effective population is computed from the observed
population explained previously). Now, the person-years lived by $B(t_{0})$
during their third year of life, (after removing person-years lost
due to deaths during third year) and assuming the deaths are uniformly
distributed over the age intervals $[2,3)$ are:

\begin{eqnarray}
\int_{t_{0}}^{t_{1}}P_{1}(s)ds & - & \frac{1}{2}\int_{t_{0}}^{t_{1}}B(s)ds+\frac{1}{2}\int_{t_{0}}^{t_{1}}D_{0}(s)ds,\label{L2}
\end{eqnarray}

The total person-years that would be lived by $B(t_{0})$ during their
first three-years of life is

\begin{eqnarray}
\frac{1}{2}\int_{t_{0}}^{t_{1}}B(s)ds+2\int_{t_{0}}^{t_{1}}P_{1}(s)ds\label{T(0) when l(3)=00003D0}
\end{eqnarray}

\begin{eqnarray*}
\end{eqnarray*}
The life expectancy of $B(t_{0})$, when $l(3)=0$ is:

\begin{eqnarray}
\frac{1}{2}+2\frac{\int_{t_{0}}^{t_{1}}P_{1}(s)ds}{\int_{t_{0}}^{t_{1}}B(s)ds}\label{LE when l(3)=00003D0}
\end{eqnarray}

Proceeding further with a similar approach, we can obtain $e(B(t_{0}))$,
the life expectancy of $B(t_{0})$ when $l(\omega)=0$ as:

\begin{eqnarray}
e(B(t_{0})) & = & \left\{ \begin{array}{cc}
\frac{3}{2}-\frac{\int_{t_{0}}^{t_{1}}D_{0}(s)ds}{\int_{t_{0}}^{t_{1}}B(s)ds}+\frac{2}{\int_{t_{0}}^{t_{1}}B(s)ds}\Sigma_{n=1}^{\frac{\omega}{2}-1}\int_{t_{0}}^{t_{1}}P_{2n}(s)ds & \mbox{if }\omega\mbox{ is even}\\
\\
\frac{1}{2}+\frac{2}{\int_{t_{0}}^{t_{1}}B(s)ds}\Sigma_{n=0}^{\frac{\omega-3}{2}}\int_{t_{0}}^{t_{1}}P_{2n+1}(s)ds & \mbox{if }\omega\mbox{ is odd}
\end{array}\right.\label{general LE}
\end{eqnarray}

\begin{figure}
\includegraphics[scale=0.8]{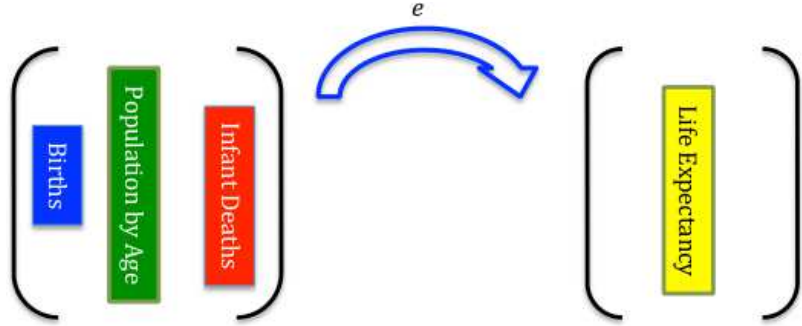}

\caption{\label{LE-limiteddata} Life expectancy with limited data. Only with
the information on births, effective population by age and infant
deaths in a year, the proposed formula will forecast the life expectancy
of newly born babies in a year. }

\end{figure}

\section*{\textbf{Numerical Examples}}

We consider an example population of some arbitrary species, whose
effective population age structures, births and infant deaths are
observed during some interval $[t_{0},t_{1})$ (see Table \ref{Table1}).
We give the computed life expectancies in Table\textbf{ \ref{Table1}}. 

\begin{table}
\caption{\label{Table1} Set of two hypothetically observed population age
structures, births, infant deaths during $[t_{0},t_{1}),$ and computed
life expectancies.}

(a) %
\begin{tabular}{|c|c|c|c|c|}
\hline 
Age & $\begin{array}{c}
\mbox{Effective}\\
\mbox{Population}
\end{array}$ & Births & $\begin{array}{c}
\mbox{Infant}\\
\mbox{Deaths}
\end{array}$ & $\begin{array}{c}
\mbox{Life}\\
\mbox{Expectancy}
\end{array}$\tabularnewline
\hline 
\hline 
0 & 10 & 12 & 1 & \textbf{4.5}\tabularnewline
\hline 
1 & 12 &  &  & \tabularnewline
\hline 
2 & 14 &  &  & \tabularnewline
\hline 
3 & 12 &  &  & \tabularnewline
\hline 
4 & 6 &  &  & \tabularnewline
\hline 
5 & 0 &  &  & \tabularnewline
\hline 
\end{tabular}

\vspace{0.5cm}

(b) %
\begin{tabular}{|c|c|c|c|c|}
\hline 
Age & $\begin{array}{c}
\mbox{Effective}\\
\mbox{Population}
\end{array}$ & Births & $\begin{array}{c}
\mbox{Infant}\\
\mbox{Deaths}
\end{array}$ & $\begin{array}{c}
\mbox{Life}\\
\mbox{Expectancy}
\end{array}$\tabularnewline
\hline 
\hline 
0 & 12 & 9 & 3 & \textbf{5.17}\tabularnewline
\hline 
1 & 16 &  &  & \tabularnewline
\hline 
2 & 18 &  &  & \tabularnewline
\hline 
3 & 12 &  &  & \tabularnewline
\hline 
4 & 0 &  &  & \tabularnewline
\hline 
 &  &  &  & \tabularnewline
\hline 
\end{tabular} 

\end{table}

We further simplify the life expectancy formula of (\ref{general LE})
based on \textbf{a} few assumptions and we obtain (\ref{general LE-simple}).
For details, see the Appendix. We tested this formula (for $\omega$
even and odd) on global population data \cite{UN-Population}. Total
population in 2010 was approximately 6916 million, and infant deaths
were 4.801 million. We have obtained $P_{\geq}(t_{0})$, the total
population size with individuals whose age is one and above by removing
the size of the population, whose age is zero, from the total population.
The adjusted $P_{\geq1}(t_{0})$ is $6756$ million. Assuming a range
of live births of 90-100 million occurred during 2010, we have calculated
that the life expectancy of cohorts born in 2010 will be between 69
- 76.5 years (when $\omega$ is even), and life expectancy for these
newly born will be 68.1 - 75.5 years (when $\omega$ is odd). In 2010,
the actual global life expectancy was 70 years. We note that the formula
in (\ref{general LE-simple}), and the assumption in (\ref{assumption 2P=00003DP})
may not be true for every population's age-structure. Interestingly
the formula results (\ref{general LE-simple}) are very close to the
life table-based standard estimates for the US and UK populations.
However, it should be noted that the formula did not work for some
populations. The total population in US in 2011 was approximately
313 million, and the total live births are approximately 4 million.
This gives us $e(B(t_{0}))=0.5+78.25=78.75$ years, whereas the actual
life expectancy for the US population for 2011 is $78.64$ years.
Similarly, the formula-based values for UK is $78.23$ years and actual
value is $80.75$ years. 

In this paper we suggest a formula for computing life expectancy of
a cohort of new born babies when it is difficult to construct a life
table based life expectancy. For the standard life table technique,
one requires information on $\int_{t_{0}}^{t_{1}}\int_{0}^{\omega}D_{i}(s)dids$,
the total deaths during $[t_{0},t_{1})$, where $\int_{0}^{\omega}D_{i}(s)di$
is the age-specific death numbers at time $s\in[t_{0},t_{1})$, and
then, traditionally compute age-specific death rates at age $i$ during
$[t_{0},t_{1})$ using, 
\begin{eqnarray}
\frac{\int_{t_{0}}^{t_{1}}D_{i}(s)ds}{P_{i}(t_{0})}.\label{eq:ASDR}
\end{eqnarray}
It is possible to obtain probability of deaths from (\ref{eq:ASDR}),
with some assumptions on the pattern of deaths within the time interval.
We compute various columns of the life table from these death probabilities
and compute life expectancy. 

The proposed formula in (\ref{general LE}) is very handy and can
be computed by non-experts with minimal computing skills. It can be
adapted by ecologists, experimental biologists, and biodemographers
where the data on populations are limited. See Figure \ref{LE-limiteddata}
for the data needed to compute life expectancy of newly born babies
in a year. It requires some degree of caution to apply the proposed
formula when sufficient death data by all age groups is available.
Our method heavily depends on the age structure of the population
at the time of data collection. Our approach needs to be explored
when populations are experiencing stable conditions given in \cite{Rao ASRS}
and also to be tested for its accuracy at different stages of demographic
transition. We still recommend to use life table methods when age-wise
data on deaths and populations are available as indicated in Figure
\ref{LT-DE Fig}.

\section*{\textbf{Acknowledgements}}

Dr. Cynthia Harper (Oxford) and Ms. Claire Edward (Kent) have helped
to correct and revise several sentences. Our sincere gratitude to
all.

\section*{\textbf{Appendix: Analysis of the Life Expectancy function}}

In general, $\int_{t_{0}}^{t_{1}}D(s)ds<\int_{t_{0}}^{t_{1}}B(s)ds$.
When $\omega$ is even, the supremum and infemum of $\left(\frac{3}{2}-\frac{\int_{t_{0}}^{t_{1}}D_{0}(s)ds}{\int_{t_{0}}^{t_{1}}B(s)ds}\right)$
are $\frac{3}{2}$ and $\frac{1}{2}.$ The contribution of the term
$\left(\frac{3}{2}-\frac{\int_{t_{0}}^{t_{1}}D_{0}(s)ds}{\int_{t_{0}}^{t_{1}}B(s)ds}\right)$
in computation of life expectancy is very minimal in comparison with
the term $\left(\frac{2}{\int_{t_{0}}^{t_{1}}B(s)ds}\Sigma_{n=1}^{\frac{\omega}{2}-1}\int_{t_{0}}^{t_{1}}P_{2n}(s)ds\right)$,
hence $e(B(t_{0}))$ can be approximated by,

\begin{eqnarray*}
e(B(t_{0})) & \approx & \frac{2}{\int_{t_{0}}^{t_{1}}B(s)ds}\Sigma_{n=1}^{\frac{\omega}{2}-1}\int_{t_{0}}^{t_{1}}P_{2n}(s)ds
\end{eqnarray*}

Similarly, when $\omega$ is even, $e(B(t_{0}))$ can be approximated
by,

\begin{eqnarray*}
e(B(t_{0})) & \approx & \frac{2}{\int_{t_{0}}^{t_{1}}B(s)ds}\Sigma_{n=1}^{\frac{\omega-3}{2}}\int_{t_{0}}^{t_{1}}P_{2n+1}(s)ds
\end{eqnarray*}

\begin{rem}
Suppose $\left(P_{n}(t_{0})\right)_{0}^{\omega}$ is an increasing,
then, we will arrive at the two inequalities (\ref{Ineq1}) and (\ref{eIneq2}).

\begin{eqnarray}
\Sigma_{n=1}^{\frac{\omega}{2}-1}\int_{t_{0}}^{t_{1}}P_{2n}(s)ds & < & \frac{1}{2}\Sigma_{n=1}^{\omega}\int_{t_{0}}^{t_{1}}P_{n}(s)ds\;\mbox{ if }\omega\mbox{ }\mbox{is even,}\label{Ineq1}\\
\nonumber \\
\Sigma_{n=1}^{\frac{\omega-3}{2}}\int_{t_{0}}^{t_{1}}P_{2n+1}(s)ds & > & \frac{1}{2}\Sigma_{n=1}^{\omega}\int_{t_{0}}^{t_{1}}P_{n}(s)ds\;\mbox{if }\omega\mbox{ is odd.}\label{eIneq2}
\end{eqnarray}
\end{rem}

$ $
\begin{rem}
In general when $\left(P_{n}(t_{0})\right)_{0}^{\omega}$ is an increasing,
without any condition on $\omega$, we can write the inequality (\ref{inequlaity(combined)})
by combining (\ref{Ineq1}) and (\ref{eIneq2}) as,

\begin{equation}
\Sigma_{n=1}^{\frac{\omega}{2}-1}\int_{t_{0}}^{t_{1}}P_{2n}(s)ds<\frac{1}{2}\Sigma_{n=1}^{\omega}\int_{t_{0}}^{t_{1}}P_{n}(s)ds<\Sigma_{n=1}^{\frac{\omega-3}{2}}\int_{t_{0}}^{t_{1}}P_{2n+1}(s)ds\label{inequlaity(combined)}
\end{equation}
\end{rem}

$ $
\begin{rem}
Suppose $\int_{t_{0}}^{t_{1}}D_{0}(s)ds=\int_{t_{0}}^{t_{1}}B(s)ds$
in (\ref{general LE}), then, the life expectancy, irrespective of
$\omega$ is even or odd, becomes,
\begin{eqnarray}
e(B(t_{0})) & = & \frac{1}{2}+\frac{2}{\int_{t_{0}}^{t_{1}}B(s)ds}\Sigma_{n=0}^{\frac{\omega-3}{2}}\int_{t_{0}}^{t_{1}}P_{2n+1}(s)ds\label{eqLE when D0=00003DB}
\end{eqnarray}
\end{rem}

$ $
\begin{rem}
When the total population aged one and above at $t_{0}$ is approximately
same as twice the sum of the populations of even single year ages
and also twice the sum of the populations of odd single year ages,
i.e. 
\begin{equation}
2\Sigma_{n=1}^{\frac{\omega}{2}-1}\int_{t_{0}}^{t_{1}}P_{2n}(s)ds\approx\int_{t_{0}}^{t_{1}}P(s)ds\approx2\Sigma_{n=0}^{\frac{\omega-3}{2}}\int_{t_{0}}^{t_{1}}P_{2n+1}(s)ds,\label{assumption 2P=00003DP}
\end{equation}
\end{rem}

then, life expectancy in (\ref{general LE}) further reduces into, 

\begin{eqnarray}
e(B(t_{0})) & = & \left\{ \begin{array}{cc}
\frac{3}{2}-\frac{\int_{t_{0}}^{t_{1}}D_{0}(s)ds}{\int_{t_{0}}^{t_{1}}B(s)ds}+\frac{\int_{t_{0}}^{t_{1}}P_{\geq1}(s)ds}{\int_{t_{0}}^{t_{1}}B(s)ds} & \mbox{if }\omega\mbox{ is even}\\
\\
\frac{1}{2}+\frac{\int_{t_{0}}^{t_{1}}P_{\geq1}(s)ds}{\int_{t_{0}}^{t_{1}}B(s)ds} & \mbox{if }\omega\mbox{ is odd}
\end{array}\right.\label{general LE-simple}
\end{eqnarray}

where $P_{\geq1}(s)$ is the effective population who are aged one
and above at time $s\in[t_{0},t_{1}).$ 
\end{document}